# Investigation of spin scattering mechanism in silicon channels of Fe/MgO/Si lateral spin valves

**Soobeom Lee[1], Naoto Yamashita[1], Yuichiro Ando[1]\*, Shinji Miwa[2], Yoshishige Suzuki[2], Hayato Koike[3] and Masashi Shiraishi[1]#**

1. Department of Electronic Science and Engineering, Kyoto University, Japan.
2. Graduate School of Engineering Science, Osaka University, Japan.
3. Technology HQ, TDK Corporation, Japan.

\* Corresponding author: Yuichiro Ando (ando@kuee.kyoto-u.ac.jp)
\# Corresponding author: Masashi Shiraishi (mshiraishi@kuee.kyoto-u.ac.jp)

**Abstract**

**The temperature evolution of spin relaxation time, $\tau_{sf}$, in degenerate silicon (Si)-based lateral spin valves is investigated by means of the Hanle effect measurements. $\tau_{sf}$ at 300 K is estimated to be 1.68 ±0.03 ns and monotonically increased with decreasing temperature down to 100 K. Below 100 K, in contrast, it shows almost a constant value of ca. 5 ns. The temperature dependence of the conductivity of the Si channel shows a similar behavior to that of the $\tau_{sf}$, i.e., monotonically increasing with decreasing temperature down to 100 K and a weak temperature dependence below 100 K. The temperature evolution of conductivity reveals that electron scattering due to magnetic impurities is negligible. A comparison between $\tau_{sf}$ and momentum scattering time reveals that the dominant spin scattering mechanism in the Si is the Elliott-Yafet mechanism, and the ratio of the momentum scattering time to the $\tau_{sf}$ attributed to nonmagnetic impurities is approximately $3.77 \times 10^{-6}$, which is more than two orders of magnitude smaller than that of copper.**

Over the past decade, spin transport phenomena in silicon (Si) have been investigated rigorously to realize the potential of Si-based spintronic devices,[1–11] which are promising candidates beyond complementary metal oxide semiconductor (CMOS) technologies.[12–15] To integrate spin functionality into the transistors, highly efficient spin injection into semiconductor channels and long-range spin transport through the semiconductor channel are essential. From this point of view, Si is expected to be a suitable channel as a result of the good spin coherence due to weak spin-orbit interactions and spatial inversion symmetry of the crystal structure. Good compatibility with mature, large-scale integration technologies also boosted concentrated research on Si-based spin transistors.[2–11] Owing to the number of research projects, the transport of spin polarized current and/or pure spin current in the Si channel has been developed several ways, e.g., by electrical,[2,16,17] optical,[4] and dynamical means.[10,11] In particular, long-range spin transport has been developed by electrical spin injection, even at room temperature (RT),[17–19] which paved the way for the spin metal-oxide-semiconductor field-effect transistor (MOSFET) operation at RT.[20,21] It has also been revealed that spin relaxation time, $\tau_{sf}$, in Si is much longer than those in other semiconductors, which consist of relatively heavy elements ($\tau_{sf}$ in germanium, Ge: 0.6 ns)[22–24] and exhibit breaking of spatial inversion symmetry in the crystallographic structure ($\tau_{sf}$ in gallium arsenide, GaAs: 0.01 ns).[25] Temperature dependence of $\tau_{sf}$ in the Si channel was also weak compared with that in GaAs, which enables stable operation of spin devices in a wide range of temperature.[26] For high performance operation of spin transistors, e.g., low power consumption, large magnetoresistance, etc., enhancement of $\tau_{sf}$ is an important issue. To enhance the $\tau_{sf}$, investigation of the spin scattering mechanism and reduction of the spin scattering centers are desired[27]. In a copper (Cu) channel, it is revealed that dilute magnetic impurities of several parts per million (ppm) dominate the spin scattering probability, resulting in a serious reduction in $\tau_{sf}$ at low temperature.[28] Enhancement of spin diffusion length, by reducing the concentration of magnetic impurity in the Cu channel, was also demonstrated at low temperature. In the Si channel, the theoretical approach of the spin relaxation mechanism, due to donor impurity, was elucidated and revealed that intervalley scattering with change in the momentum axis in the $k$-space ($f$-process) caused by short-range scattering off the central-cell potential of the impurities is the dominant factor of the spin relaxation.[29] The contribution of the electron-phonon scattering to the spin relaxation was also investigated. In this case, an intervalley $f$-process, due to scattering with shortwave Σ-axis phonon, was a main contributor of spin scattering.[30–32] Although theoretical approaches have successfully revealed the underlying physics of the spin scattering phenomena in Si, comparisons between theoretical and experimental $\tau_{sf}$ have been limited at low temperature and specific measurements, such as electron paramagnetic resonance, primarily because of the difficulties in experimentally demonstrating electrical spin transport in a wide range of temperature and impurity concentrations and a variety of impurity species.

In this paper, we investigate the spin scattering mechanism and spin scattering probability of each momentum scattering caused by impurity and phonon in the Si channel in a wide temperature range from 4.2 to 300 K. The Si-based lateral spin valves equipped with Fe/MgO ferromagnetic spin injectors and detectors show clear spin transport signals in the 4.2 K to 300 K temperature range. $\tau_{sf}$ is monotonically increased with decreasing the temperature to 100 K. In contrast, it shows almost a constant value below 100 K. Temperature dependence of the conductivity of the Si channel shows a similar behavior to that of the $\tau_{sf}$ i.e., monotonically increasing with decreasing temperature down to 100 K and a weak temperature dependence below 100 K. The maximum conductivity is obtained at the lowest temperature, indicating that electron scattering due to magnetic impurities is negligible, even though Fe/MgO bilayers in the channel region were removed by argon ion ($Ar^+$) milling, which might induce Fe impurities in the Si channel. A comparison between $\tau_{sf}$ and momentum scattering time reveals that the dominant spin scattering mechanism in the Si is the Elliott-Yafet mechanism, and the ratio of the momentum scattering time to the $\tau_{sf}$ attributed to nonmagnetic impurities is approximately $3.77 \times 10^{-6}$, which is more than two orders of magnitude smaller than that of the Cu channel.[28,33]

The device structure of the Si-based lateral spin valves (LSVs) in this study is schematically shown in Fig. 1(a). The Si-based LSV was fabricated on a silicon-on-insulator substrate with 100-nm-thick Si(100) / 200-nm-thick $SiO_2$ / bulk Si(100). The Si channel layer was fabricated by ion implantation of phosphorous (P). Dopant concentration, as measured by secondary ion mass spectrometry, was approximately $N_D = 5\times10^{19}$ $cm^{-3}$. Through a 4-terminal method, the conductivity of the degenerate Si channel was determined to be $5.61\times10^4$ $\Omega^{-1}m^{-1}$ at RT. Pd (3 nm) / Fe (13 nm) / MgO (0.8 nm); layers were grown on the Si channel by molecular beam epitaxy. Ferromagnetic metal electrodes were fabricated by electron beam lithography and $Ar^+$ ion milling. The channel region between the ferromagnetic contacts was over-etched to a depth of approximately 15 nm. Magnetoresistance measurements and Hanle effect measurements were carried out in a liquid helium flow cryostat system using a commercial source measure unit and digital multimeter.

Typical results of nonlocal four-terminal (NL-4T) measurements[34] measured at 4.2 and 300 K are shown in Fig. 2(a). DC current was 3 mA and in-plane magnetic fields were applied along ±y direction in Fig 1(a). Clear rectangular hysteresis signals were obtained both at 4.2 and 300 K, indicating successful spin transport through the Si channel layer. The magnitude of spin accumulation signals was estimated to be approximately 0.3 and 0.03 mV at 4.2 and 300 K, respectively. Hanle effect signals in the NL-4T scheme under parallel configuration at various temperatures, where magnetic fields were applied along perpendicular to the film plane (±z direction in Fig. 1(b)), are displayed in Fig. 2(b). Clear Hanle effect signals were obtained for all temperatures. To estimate $\tau_{sf}$ in the degenerate Si, we used the following fitting function[35], which is an analytical solution of the spin drift diffusion equation:

$$\frac{V_{NL}(B)}{I} = \frac{P^2 \lambda_N}{2\sigma A}(1+\omega^2\tau_{sf}^2)^{-\frac{1}{4}} \exp\left\{-\frac{d}{\lambda_N}\sqrt{\frac{\sqrt{1+\omega^2\tau_{sf}^2}+1}{2}}\right\} \left\{\cos\left(\frac{\arctan(\omega\tau_{sf})}{2} + \frac{d}{\lambda_N}\sqrt{\frac{\sqrt{1+\omega^2\tau_{sf}^2}-1}{2}}\right)\right\}, \quad ...(1)$$

where $P$ is the spin polarization of the injected current, $\sigma$ is the conductivity of the silicon channel, $A$ is the cross-sectional area of the channel, $d$ is the center to center gap distance between ferromagnetic electrodes, $\omega = g\mu_B B/\hbar$ is the Larmor frequency, $g$ is the $g$-factor for the electrons ($g = 2$ in this study), $\mu_B$ is the Bohr magneton, and $\hbar$ is the Dirac constant. The spin diffusion length $\lambda_N$ is given by $\lambda_N = (D\tau_{sf})^{0.5}$, where $D$ is the spin diffusion constant. In the analyses, $A=9.35\times10^{-13}$ m$^2$ and $d=1.65$ μm were fixed, and $D$, $\tau_{sf}$, and $P$ were derived from the fitting using Eq. (1). Typical fitting curves are shown in solid curves in Fig. 2(b). The fitting curves nicely reproduce the experimental results. $\tau_{sf}$ and $\lambda_N$ were estimated to be 5.14 ± 0.07 and 1.93 ± 0.02 μm at 4 K, 1.68±0.03 ns and 1.08 ± 0.02 μm at 300 K, respectively, consistent with previous studies[17]. The temperature dependence of $\tau_{sf}$ obtained from Eq. (1) and the magnitude of the Hanle signals $\Delta V_S$ are displayed in Fig. 3(a) and 3(b), respectively. A monotonic increase of $\tau_{sf}$ with decreasing temperature and plateau below 100 K within the experimental uncertainty were obtained. $\Delta V_S$ also shows a monotonic increase with decreasing temperature followed by a plateau behavior below 100 K. Reduction in $\tau_{sf}$ and $\Delta V_S$ at low temperature reported in the Cu channel, including several ppm Fe atoms or CuFe alloys, was not observed.[28,36] Figure 3(c) shows the temperature dependence of resistivity, $\rho_{Si}$, of the Si channel using the conventional four terminal method, whose current-voltage scheme is shown in Fig. 1(c). The $\rho_{Si}$ at 300 K was estimated to be ca. 17.8 Ωμm and monotonically decreased with decreasing temperature, indicating a degenerate Si channel. Although $\rho_{Si}$ shows almost a constant value below 100 K, the minimum value was obtained at the lowest temperature, indicating that the Kondo effect is negligibly small in our Si-based LSVs, which is consistent with the temperature evolution of $\tau_{sf}$ and $\Delta V_S$. In the device fabrication procedure, Fe/MgO contacts were fabricated by using Ar$^+$ ion milling, which might induce Fe impurities in Si. However, a simple Monte Carlo simulation using

TRIM[37] revealed that the Si region with 1 ppm Fe impurity was limited and approximately within 4 nm from the surface. Over-etching of the Si channel at approximately 15 nm successfully removes the unwanted Si layer. Therefore, absence of the Kondo effect in our Si-LSVs is reasonable. From Matthiessen's rule, $\tau_{sf}$ is expressed as:

$$\frac{1}{\tau_{sf}} = \frac{\epsilon_i}{\tau_e^i} + \frac{\epsilon_p}{\tau_e^p} + \frac{\epsilon_m}{\tau_e^m}, \quad ...(2)$$

where $\tau_e^i$, $\tau_e^p$, and $\tau_e^m$ are the momentum scattering time of the electrons due to nonmagnetic impurities, phonons, and magnetic impurities, respectively. $\epsilon_\alpha (\alpha = i, p, m)$ is the spin flip probability of each momentum scattering. The $\frac{\epsilon_m}{\tau_e^m}$ term is expected to be zero in our Si LSVs, as discussed in Fig. 3. Since the phonon scattering is suppressed at low temperature (4.2 K), momentum scattering due to nonmagnetic impurities is expected to be dominant at 4.2 K. At low temperature, electron-electron scattering should be considered. If the electron-electron scattering mechanism becomes dominant, a linear relationship between conductivity and $T^{0.5}$ is expected[38]. However, such a linear region was not recognized in our Si channel even at low temperature as shown in the inset of Fig. 3(d), where two different devices i.e., an Fe/MgO/Si LSV and a Si Hall bar with nonmagnetic AuSb contacts were measured. In contrast, temperature dependence of the conductivity was nicely reproduced with $\left(\alpha/T^{-\frac{2}{3}} + \beta\right)^{-1}$, where α and β are the free parameters, indicating that dominant scattering mechanism of electron at low temperature in not the electron-electron scattering but the impurity scattering. It is also noted that the similar temperature dependence of conductivity between two different devices supports our claim that $\frac{\epsilon_m}{\tau_e^m}$ term is negligible because the Hall bar device has no opportunities of magnetic impurities during the device fabrication process. Quantum interference effects (QIE) should also be considered at low temperature[39,40]. Although the weak localization was actually measured below 30 K as shown in Fig. 4(a), correction of the Drude model due to the QIE was estimated to be less than 0.4% even at 4 K. Thus the QIE is also negligible in our device. The dimensionless transport parameter $k_F l$, where $k_F$ is the Fermi wavevector and $l$ is the mean free path in the Si channel, much larger than 1 also support our claim (see Fig. 4(b)). Therefore, hereafter we discuss the spin scattering mechanism based on the Drude model. From the Drude model, $\rho_{Si}$ is expressed as:

$$\rho_{Si} = \frac{m^*}{e^2 n \tau_e}, \quad ...(3)$$

where $e$, $n$, and $m^*$ are the elementary charge, carrier concentration, and effective mass of electrons, respectively. Using $\tau_{sf}$ =5.14 ns obtained from Hanle measurements at 4.2 K, $m^*$=2.37

$\times 10^{-31}$ kg and $\tau_e^i=1.93\times10^{-5}$ ns obtained from Fig. 3(b) and Eq. (3); $\epsilon_i$ was estimated to be $3.77\times10^{-6}$.

Hereafter, we roughly estimate spin scattering probability at 300 K. Temperature dependence of $\tau_{sf}$ is expressed as follows:

$$\frac{1}{\tau_{sf}(T)} = \frac{\epsilon_i}{\tau_e^i(T)} + \frac{\epsilon_p}{\tau_e^p(T)}, \qquad ...(4)$$

where $\tau_e^i(T)$ and $\tau_e^p(T)$ are generally temperature dependent parameters in nondegenerate semiconductors. In contrast, in the degenerate Si, the temperature dependence of $\tau_e^i(T)$ is negligibly small because of electron screening of ionized impurities.[41] Therefore, we used a constant $\tau_e^i(T)=1.93\times10^{-5}$ ns. Similar to the Cu channel, we assumed temperature independent $\epsilon_i$ and $\epsilon_p$.[33] Using $\rho_{Si}$ and $\tau_{sf}$ at RT obtained experimentally, $\epsilon_p$ was estimated to be $9.02\times10^{-6}$. Finally, the temperature dependence of $\tau_{sf}$ was calculated by using obtained $\epsilon_p$ and $\epsilon_i$ and the temperature dependence of $\rho_{Si}$. Assuming a constant $\tau_e^i(T)$ and the Drude model, $\tau_{sf}$ was calculated from Eq. (4). The result is shown in Fig. 4. Even for such a rough assumption, the calculated $\tau_{sf}$ nicely reproduces the experimentally obtained one from the Hanle measurements, indicating that the obtained parameters are credible. $\epsilon_i$ and $\epsilon_p$ of less than 1 indicate that the dominant spin scattering mechanism in the Si is the Elliott-Yafet mechanism, and these values are approximately two-orders of magnitude smaller than those in the Cu channel, indicating good spin coherence in Si. It is also suggested that spin scattering time due to phonon scattering becomes comparable to that of the nonmagnetic impurity scattering of approximately 190 K in our Si channel and dominant at 300 K, which is in good agreement with the theoretical results.[29] This is also consistent with $\tau_{sf}$ in nondegenerate Si ($N_D=2\times10^{18}$ cm$^{-3}$), where comparable $\tau_{sf}$ (1~3 ns) with that in the degenerate Si was obtained at 300 K, regardless of more than an order of difference in impurity concentration. A strained Si channel might reduce the spin scattering probability induced by phonon scattering and realizes further enhancement of spin lifetime at room temperature.[42]

In summary, the spin relaxation mechanism and spin scattering probability of each momentum scattering in degenerate Si-based LSVs have been investigated in a wide temperature range from 4.2 to 300 K. The dominant spin scattering mechanism was the Elliott-Yafet mechanism, and spin scattering probabilities, due to impurity and phonon scattering, were estimated to be $3.77\times10^{-6}$ and $9.02\times10^{-6}$, respectively, which are approximately two-orders of magnitude smaller than those in the Cu channel. The dominant spin scattering was impurity scattering at 4.2 K and phonon scattering at 300 K.

**Acknowledgements**

This research was supported in part by a Grant-in-Aid for Scientific Research from the Ministry of Education, Culture, Sports, Science and Technology (MEXT) of Japan (Innovative Area "Nano Spin Conversion Science"No. 26103002 and No. 26103003, Grant-in-Aid for Scientific Research (A) No. 25246019, Grant-in-Aid for Scientific Research (S) "Semiconductor Spincurrentronics" No. 16H06330 and Grant-in-Aid for Young Scientists (A) No. 16H06089).

**Additional information**

The authors declare no competing financial interests.

**Figure captions**

**Figure 1**

(a) Schematic of the silicon (Si)-based lateral spin valve (LSV). Two Fe/MgO electrodes were fabricated as a spin injector and detector. Center-to-center distance between two electrodes was 1.65 μm. Current-voltage configurations for measurements of (b) the Hanle effect signal and (c) channel resistivity. In the Hanle effect measurements, magnetic field $B_z$ was applied along a perpendicular direction to the film plane.

**Figure 2**

(a) Nonlocal magnetoresistance curves measured at 300 (upper panel) and 4.2 K (lower panel), where the in-plane magnetic field was applied along the long axis of ferromagnetic electrodes (y-direction in Fig. 1(a)). Current-voltage configuration was the same with that shown in Fig. 1(b). (b) Nonlocal Hanle effect curves obtained at 4.2, 50, 100, 200, and 300 K, where difference in the nonlocal voltage between parallel and antiparallel configurations are plotted. A DC current of 3 mA and perpendicular magnetic field were applied. Black solid curves are fitting curve using Eq. (1).

**Figure 3**

(a) Temperature dependence of the spin relaxation time $\tau_{sf}$ obtained from fitting of the nonlocal Hanle curves. (d) The temperature dependence of magnitude of the Hanle signals $\Delta V_S$. (c) Temperature dependence of electrical resistivity of the Si channel. Inset shows $\Delta\rho$ i.e., difference between resistivity and minimum resistivity $\rho_0$. (d) Temperature dependence of conductivity of the Si channel. A Hall bar shaped device with nonmagnetic AuSb electrodes was fabricated. Data of the circles and inverted triangles are conductivity of Si of the Fe/MgO/Si lateral spin valves and a Hall bar device, respectively. Solid lines are fitting curve with $\left(\alpha/T^{-\frac{2}{3}} + \beta\right)^{-1}$, where α and β are the free parameters. The inset shows conductivity of the Si channel as a function of $T^{0.5}$.

**Figure 4**

(a) Magnetoresistance of the Si channel at various temperatures. A DC charge current of 3 mA and perpendicular magnetic field were applied. (b) Temperature dependence of $k_F \cdot l$ in the Si channel, where $k_F$ is the Fermi wavevector and $l$ is the mean free path.

**Figure 5**

Comparison between theoretically and experimentally estimated $\tau_{sf}$. Theoretical $\tau_{sf}$ was calculated from resistivity and temperature independent $\varepsilon_i$ and $\varepsilon_p$.

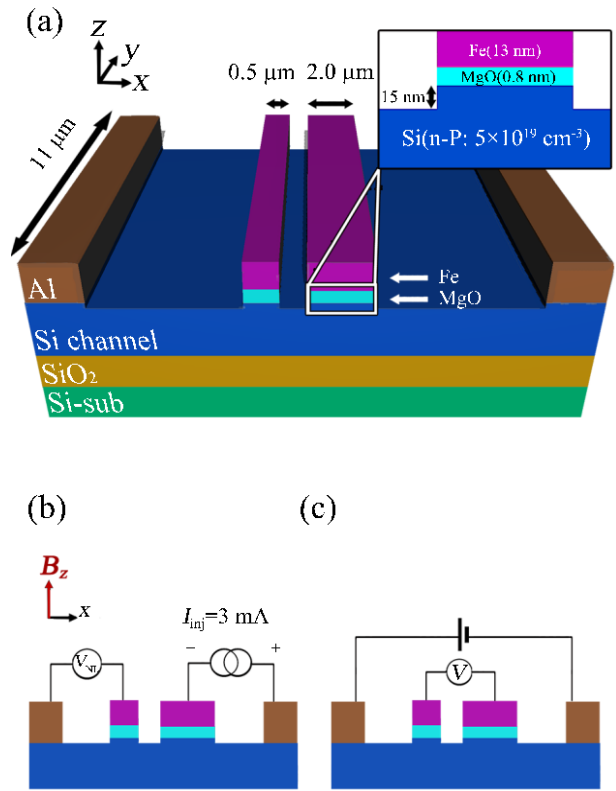

Fig. 1 Lee et al.,

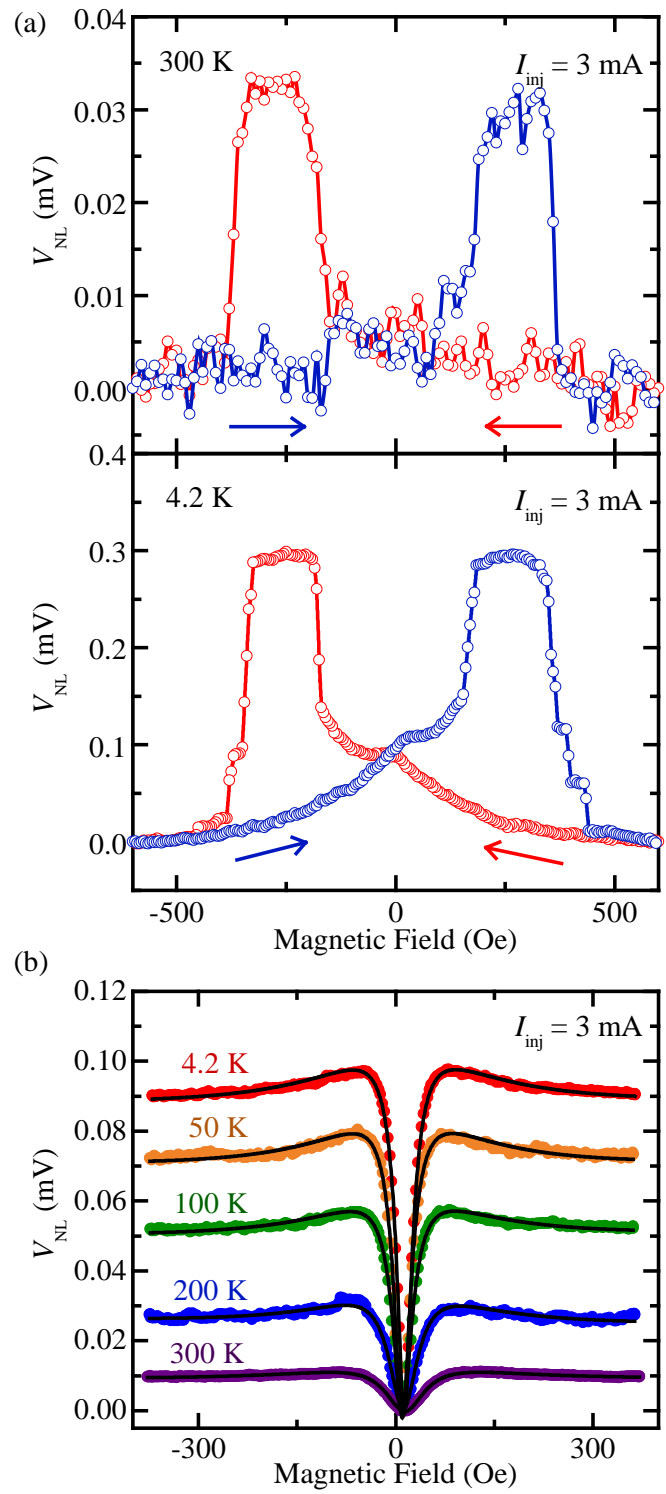

Fig. 2 Lee et al.,

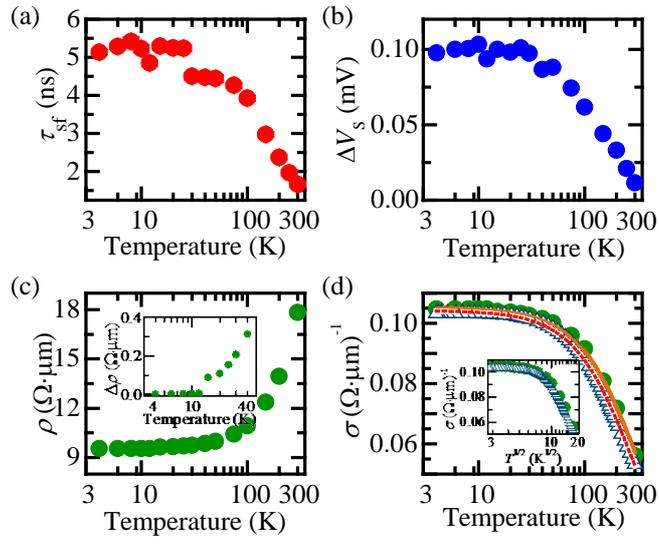

Fig. 3 Lee et al.,

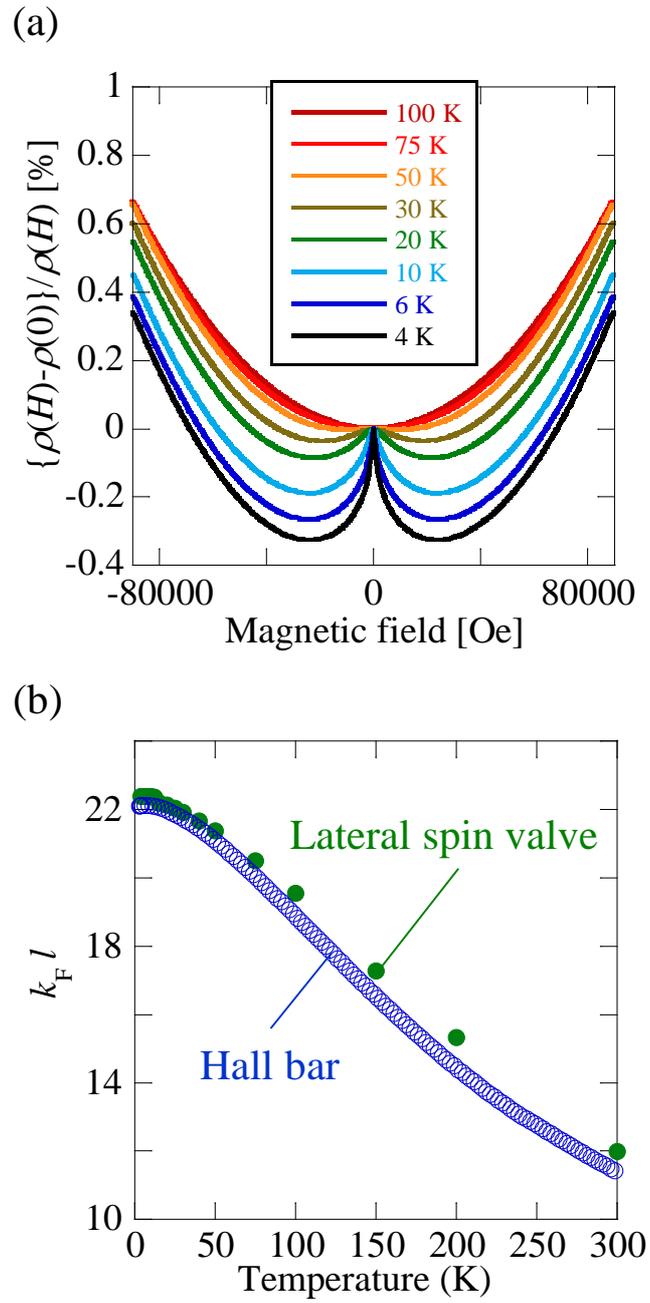

Fig. 4 Lee et al.,

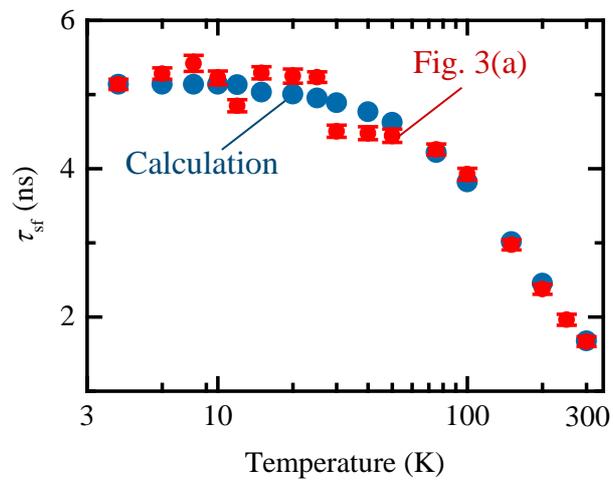

Fig. 5 Lee et al.,